# Pline: automatic generation of modern web interfaces for command-line programs

Andres Veidenberg, Ari Löytynoja

Institute of Biotechnology, University of Helsinki, Helsinki, Finland

## Abstract

**Motivation:** Bioinformatics software often lacks graphical user interfaces (GUIs), which can limit its adoption by non-technical members of the scientific community. Web interfaces are a common alternative for building cross-platform GUIs, but their potential is underutilized: web interfaces for command-line tools rarely take advantage of the level of interactivity expected of modern web applications and are rarely usable offline.

**Results:** Here we present Pline: a lightweight framework that uses program descriptions and web standards to generate dynamic GUIs for command-line programs. With Pline, cross-platform graphical interfaces are easy to create and maintain, fostering user-friendly software in science.

**Availability and Implementation:** Pline is cross-platform, open-source software. Documentation, example plugins and source code is freely available from http://wasabiapp.org/pline.

**Contact:** andres.veidenberg@helsinki.fi

## 1 Introduction

Graphical user interfaces (GUI) are an essential part of modern software, providing users with an intuitive method to access all of the functionality offered by a program. A well-designed GUI guides users by, for example, displaying relevant actions, adapting to user input, providing info tooltips and other visual cues. Software developed for research purposes, however, rarely includes a GUI, relying solely on the command line interface (CLI). The CLI is appropriate for advanced users, who can then quickly integrate new software into existing pipelines, but it comes with a steep learning curve (that may never be overcome) for non-technical users. CLIs are sensitive to typing errors, requiring users to remember which command-line options need to be included to run a program effectively. Moreover, command-line tools generally have a single, well-defined function and therefore require additional programs to view or post-process the output, further complicating their usage. These issues may appear trivial, but, in practice, they dramatically limit the number of potential users to those who are already comfortable using the command-line.

GUIs make programs more user-friendly and ease the adoption of novel computational tools. While the inclusion of GUIs is in the interest of both developers and users, there is little incentive to code a native graphical user interface. Many journals are dismissive of the scientific contribution offered by more usable software and development teams in academia tend to be constrained in terms of members (Mangul *et al.*, 2019). A common compromise is to set up an online service where a web interface is used to launch a CLI program on a remote server. This provides cross-platform, installation-free access to the software, but has many disadvantages: (i) it requires a web server, necessitating additional setup, programming and maintenance; (ii) it cannot be used offline; and (iii) it is limited by the fact that users need to share the available network and CPU resources. Furthermore, such web interfaces tend to be based on basic HTML forms with little interactivity to guide the users. While it is possible to develop more sophisticated web interfaces, this requires extensive knowledge of web technologies like CSS (Cascading Style Sheets) and Javascript, and can take a long time to develop.

Although it is considerably easier to implement a web service than a native GUI, integrating multiple programs remains a challenge. Both the web interface and the server side code needs to be built for a specific CLI program. Moreover, most of the user input processing and related code typically resides on the server that is hidden from the users. This causes redundancy, because the web interfaces cannot be reused by third-party developers for modification and therefore need to be created from scratch. The issue could be addressed by standardizing communication between a graphical interface and its program. Some standard specifications have been developed, like the Common Workflow Language (https://commonwl.org) or the Galaxy tools XML (Afgan *et al.*, 2018), that define how to describe a CLI program in a text file. Scientific workflow management systems (e.g. Taverna (Wolstencroft *et al.*, 2013) or CWL implementations like Arvados (https://arvados.org)) utilize this information to integrate external programs and, in some cases (e.g. Galaxy), also for creating the graphical interface elements. However, these standards are optimized for building pipelines and therefore omit GUI-specific instructions. In addition, setting up and running a workflow management system together with its environment (e.g. a dedicated web server, system container or virtual machine) adds unnecessary overhead when used only for creating a GUI for a CLI program.

Here, we introduce Pline (*"Plugin interface language"*): a specification for describing command line programs and their interfaces, and a lightweight framework that uses the program descriptions to build interactive graphical interfaces. By utilizing standardization and web technologies, Pline allows for creation of cross-platform GUIs without programming, considerably lowering the bar to develop user-friendly software in science.

## 2 Methods

On a technical level, a CLI program powered by a Pline interface consists of three parts: the plugin files (the program and its description in JSON format), the interface generator (a Javascript library) and the server module (a Python script). Bundled together, they form a fully functional GUI program that can be used as a standalone desktop application, as a dedicated web service, or as part of an existing web page.

## 2.1 Pline plugin API

Pline is an application programming interface (API) for writing plugins, and its implementation as a web application. A Pline web app can include one or more plugins, i.e. CLI programs and their description files. Description files are written in plain text using the API specification and are used to detail how to launch the associated program and draw its interface.

Pline plugin descriptions specify a list of command-line arguments available for a given CLI program. The information is written in JSON notation (https://www.json.org) – a common format that structures textual data with nested brackets, designed for easy reading and writing for both humans and programs. The Pline API represents interfaces using the JSON format where the data (a set of properties with values) define the underlying functionality and the structure (the order and nesting of the data items) reflects the placement of resulting interface elements. Since most of the properties are optional, basic interfaces are quick to construct (see example in Figure 1). However, the plugin description is extendable with advanced inputs and properties for sophisticated GUIs. Pline supports both simple input types like text, files, on/off flags or selection lists, and advanced ones that merge or modify values from linked inputs (see Figure 2 for an advanced case).

Input arguments are often related. A CLI program may require one or more sets of arguments where the list of compulsory or optional arguments, their values or the effect of those values is defined by the user input for a set of other arguments. In the Pline JSON format, the network of linked inputs is described by setting rules for the input properties that support it. For example, instead of a fixed default value for an input, a rule can derive the value from another input. These rules are written as conditionals – english-like if-else sentences or Javascript statements where the action of the rule is defined by the property that the rule is attached to. In addition to setting the default value, Pline supports conditionals for dynamically formatting or fixing an input value, for specifying an output file name and for enabling or disabling input elements and element groups. Together with other advanced features like input filters, error messages and merged values, the Pline API allows for describing even highly complex command-line interfaces. To customize the resulting interface, the API also specifies properties for adding icons, labels, documentation, HTML markup and CSS styling.

## 2.2 Implementation

Pline includes an interface generator that implements the JSON specification and translates the program description into a graphical user interface. The generator is written in Javascript, which runs natively inside any modern web browser, and is used by including it in a webpage as a library. The library takes a Pline plugin description as input (given as raw JSON text or a URL), parses the information into an internal data model, and outputs a graphical interface that can be placed into any container element in a web page. In practice, the process involves just two Javascript commands (`addPlugin()` for the plugin import, `plugin.draw()` for interface drawing) and supports multiple input programs and output interfaces in a single web page.

A Pline-generated GUI is not static – the HTML interface is bound to an internal data model and event listeners that enforce the dependency rules between the program parameters and adjust the interface according to user interactions. User input is tracked in real-time: as soon as a tick-box is clicked or a number is typed, the interface updates accordingly, e.g. by hiding, revealing, or changing the values of all the linked input elements. The conditionals in the Pline JSON therefore provide a quick way to construct sophisticated interfaces that hide invalid inputs and guide the user through program configuration options. In addition to generating standalone GUIs, Pline can chain multiple interfaces together, forming a pipeline – a set of commands executed in succession. The information about input and output files in the program description is used to control the data flow between the pipeline steps. The current state of a single interface or a full pipeline can be stored to a file and distributed as a reusable Pline pipeline with pre-filled input values.

A graphical interface generated with Pline records user input and forms a complete terminal command for launching the CLI program or pipeline together with user-supplied input files. Since the web browser security sandbox prevents direct command-line access, the program launch data is passed on to a backend server for execution. Pline includes a lightweight python script that acts as a server module to launch the commands, either on a local computer or over the web. The Pline server accepts the command data sent by the interface via an HTTP POST request, sanitizes the input, and manages the execution process. It also supports HTTP requests to send execution status updates back to the interface, pause, cancel or resume running pipelines, or send email notifications after a command or pipeline has finished.

Pline interfaces are designed to be installation-free and work across many different operating systems. The JSON program description files are platform-agnostic, the interface generator runs on any device with a modern web browser (including mobile devices), and the server module supports both Python 2.7 and 3 environments (which is preinstalled on most MacOS and Linux systems). However, command-line executables are compiled to run on a specific operating system, so a Pline plugin should include an executable for each target system.

## 3 Results

### 3.1 Availability

The JSON API specification for writing Pline-compatible program descriptions is available at http://wasabiapp.org/pline. The webpage also contains the documentation, the Pline implementation files for converting CLI programs to graphical applications, and a repository of example plugins.

### 3.2 Example plugins

New program interfaces are added to Pline by supplying the corresponding JSON description files, either by copying them into the designated plugin directory (when using the Pline server module), or by feeding the data directly into the Pline interface generator using a special Javascript command. In principle, an interface can be generated for any command-line executable, including installed programs that are available system-wide. However, the JSON description is written for a specific version of a program and it is recommended that the matching binary (preferably with copies for different operating systems) is distributed together with the description file, forming a plugin bundle. The Pline homepage includes several example plugins that can be used as templates for writing third-party plugins.

A basic interface can be constructed from a few pieces of information in a program description file. The only compulsory data fields are the executable name and a type and/or name property for each of the input arguments. A simple example is shown in Figure 1. Here, the JSON description specifies a python script that expects an input file (as a positional argument) and an optional boolean flag `"--count"` (a named argument). Pline translates this information into a file input element and a checkbox element. In addition, the interface header displays the program description and provides an option to name the program execution session, as well as to save or restore sets of user input values. The `"Run"` button produces the program command that is sent together with the input file to the server module for execution. In case the input file is not provided, the error message `"Input file missing!"` is displayed instead (as specified by the "`required`" property).

```
{
    "program": "remove_gaps.py",
    "name": "Gaps remover",
    "desc": "Trims gaps-only sites from the input sequence alignment",
    "outfile": "output.fa",
    "options": [
        {"file": "", "required": "Input file missing!"},
        {"checkbox": "--count", "title": "Count sequences"}
    ]
}
```

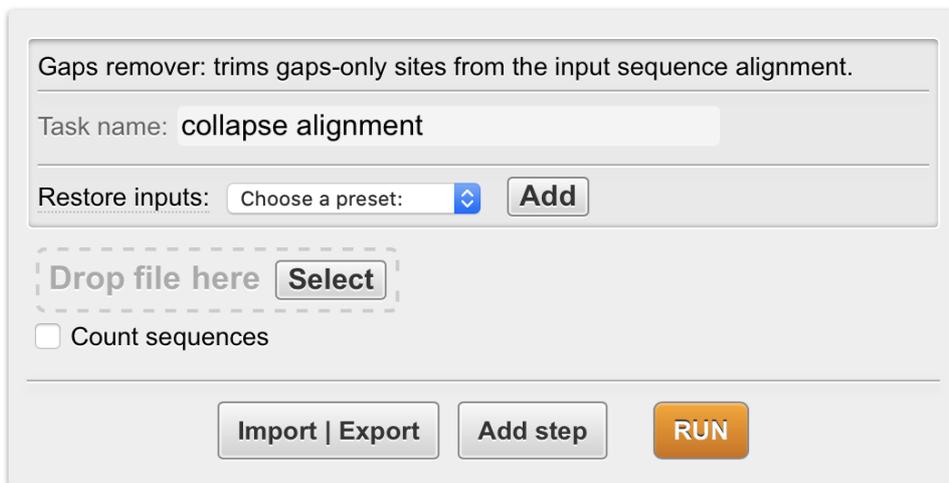

**Figure 1.** An example of a JSON-formatted description of a Pline plugin (top), and the resulting graphical interface, generated by the Pline javascript library (bottom).

With a longer list of input arguments and advanced properties like conditionals, Pline's interface rendering and command generation process is scalable to more complex programs. Examples of sophisticated interfaces can be found from the Wasabi web app (Veidenberg *et al.*, 2016) that uses Pline to draw interfaces for external tools into its graphical analysis environment.

### 3.3 Integration into Wasabi

As Pline consists of human-readable plain text files, the system is easy to customize. The appearance of the graphical interfaces can be changed by editing the style definitions in the CSS file and the Pline javascript library can be extended with custom functions to modify any step in the interface generation process. This flexibility is especially useful for integrating Pline into existing web pages.

As an example of extensive integration, we added Pline to Wasabi, a web-based environment for evolutionary sequence analyses (http://wasabiapp.org). With Pline, we were able to integrate external analysis programs into Wasabi's graphical interface as plugins without having to write program-specific interfaces and related code from scratch. Wasabi-specific features were added to the Pline interface generator as extensions. For example, an additional function in the plugin registration step makes a pop-up menu showing all the available plugins in Wasabi. Other additions automatically convert user-supplied files to the correct format, show the status updates of running programs in the Wasabi menu bar, and collects the resulting output files in an analysis database. The extensions are available as open-source software at http://wasabiapp.org.

Since Wasabi is designed for evolutionary sequence analysis, the list of existing plugins include tools for related tasks: PRANK (Löytynoja and Goldman, 2005), PAGAN (Löytynoja *et al.*, 2012) and MAFFT (Katoh and Standley, 2013) for multiple sequence alignment; FastTree (Price *et al.*, 2010) for phylogenetic inference; CodeML (Yang, 2007) for tests of positive selection. Out of these examples, CodeML has the most complex interface and serves as a comprehensive example that utilizes a majority of the options available in the Pline API. The CodeML interface in Wasabi is shown in Figure 2. The CodeML plugin JSON (and therefore its interface) includes multiple presets – stored sets of pre-filled argument values – that are useful for running common configurations of selection models and related parameters. When users select a preset, the interface hides or reveals the relevant inputs, fills these with default values and enables the corresponding models from a set of tick boxes. When the user modifies an input that is part of the selected preset, the interface checks for dependencies and changes the preset selection as the combination of the input values no longer matches the initial preset. As an example of proxy inputs, the set of model selection tick boxes are converted to their corresponding command-line form as a single argument that consists of a string of numbers representing the selected model. As CodeML takes the input arguments through a configuration file, the `configfile` and `valuesep` properties in the JSON data instruct Pline to store the argument values as a newline-delimited text file and to launch the program with the file path as its only input argument.

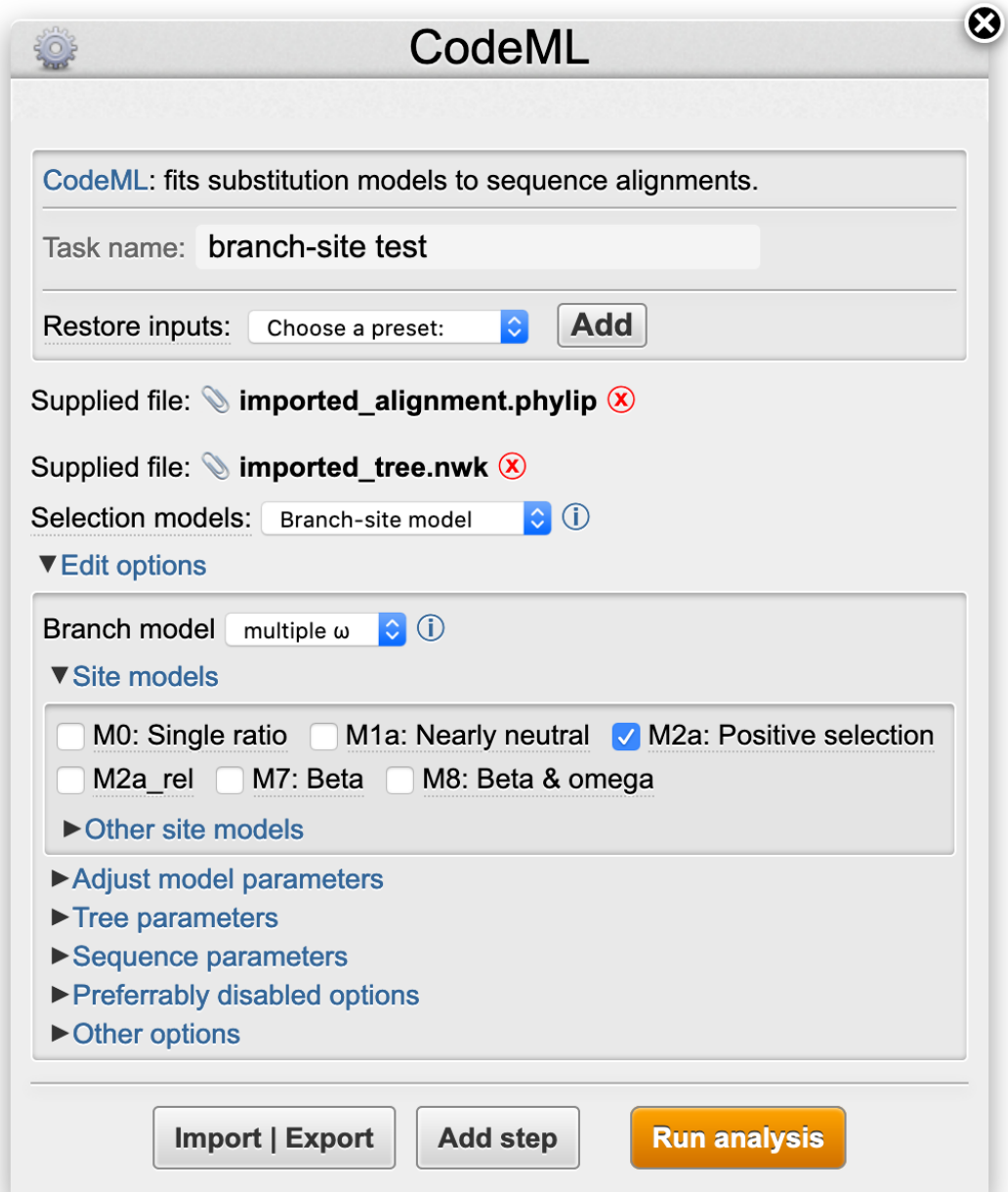

**Figure 2.** Pline interface for the CodeML plugin, rendered inside the Wasabi interface window. The input files (marked with the paperclip icon) have been automatically supplied by the Wasabi environment. Info icons and underlined text show relevant tooltips on mouseover.

### 3.4 Example pipeline

The "Add a step" button at the bottom of the Pline interface (Figure 2) allows users to quickly build a pipeline of commands by picking programs from the list of imported plugins. The plugin interfaces are drawn as a stack of collapsible sections, numbered by the order of execution. When all the inputs have been filled as needed, the resulting pipeline can be stored in a JSON file with the "Import/Export" button. The button also allows for restoring pipelines from existing JSON files to be rerun (e.g. with different input files), providing a convenient system for reusable and distributable graphical pipelines for command-line programs.

Pline pipelines with pre-filled inputs can be published in a similar manner to plugins as importable JSON files or standalone Pline packages in a public repository. For example, the repository on the Pline website includes a common analysis pipeline that maps short sequencing reads to a reference genome. It consists of four steps, starting with BWA-MEM (Li and Durbin, 2009) for mapping reads, followed by a series of Samtools commands (Li *et al.*, 2009) for converting, sorting and indexing the sequencing reads pileup (see Figure 3). The downloadable plugin package includes all the files needed for standalone execution, including JSON descriptions of the pipeline and plugins, the Pline interface generator and example input sequence data. Launching the server script (`Pline_server.py`) will subsequently open a web browser window with the graphical pipeline interface. Each of the pipeline sections can be expanded with a mouse click to examine and modify the pre-filled inputs as needed. For example, the merged sections for the first two steps indicate that these programs are launched as a piped command, where the output from the first program (*BWA-MEM*) is directly streamed as input for the second one (*Samtools view*). By changing the file input selection in the *Samtools view* interface from "pipe" to "standard output", the commands will be separated and the intermediate output file will be instead written to disk. After the included example files have been dragged to their respective filedrop areas in the *BWA-MEM* interface, the pipeline can be run.

Although the web page container in the example pipeline package includes a minimal interface for displaying the status and the results of the pipeline after it has been launched, this functionality is outside the scope of the Pline interface generator. Instead, Pline offers a framework for web developers and scientists to integrate graphical interfaces for command-line programs to websites with very low effort, especially when the needed plugin descriptions are already available. However, the example code for the post-launch interface in the plugin package and Wasabi are open-source and can be used as-is or modified for custom integration. When Pline plugin is used as a standalone interface in desktop application form, the results retrieval interface is not needed as the files are directly accessible in the work directories specified by the Pline server configuration file.

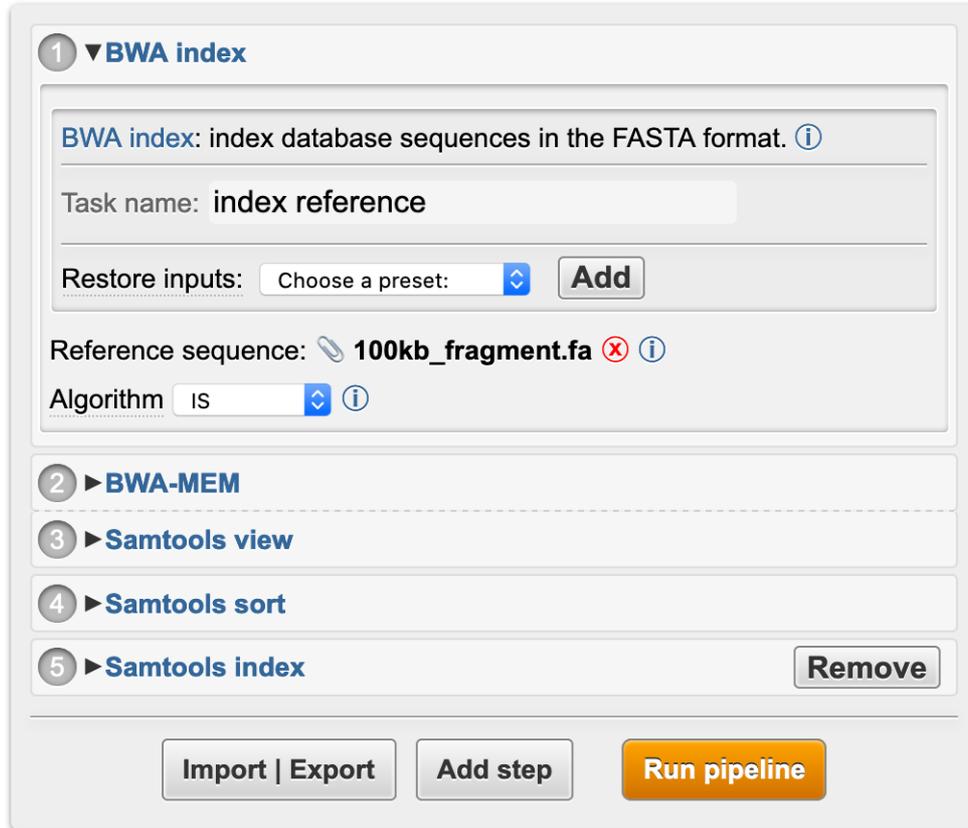

**Figure 3.** Graphical interface of the example pipeline, generated after importing its JSON file. The first pipeline step (BWA index) has been expanded showing the file input that has been automatically filled with example data.

## 4 Discussion

The lack of graphical user interfaces is a common obstacle for published computational tools becoming more widely adopted by the scientific community. The importance of user-friendly analysis software is indicated by the popularity of graphical environments for command-line programs used in bioinformatics. For example, the commercially licensed Geneious software, which provides common command-line tools in a user-friendly analysis environment, advertises itself as the most cited software in molecular biology and sequence analysis (https://www.geneious.com). Well-known open-source alternatives to Geneious include stand-alone analysis software (e.g. UGENE (Okonechnikov *et al.*, 2012)), specialized web applications (e.g. qPortal (Mohr *et al.*, 2018) or cBio Cancer Genomics Portal (Cerami *et al.*, 2012)) and workflow management systems that have a broad application field but provide only a limited graphical interface (e.g. Galaxy (Afgan *et al.*, 2018)) or are designed for CLI only (e.g. CWL implementations (https://commonwl.org)).

Pline uses standardized program descriptions to automatically generate graphical interfaces. In addition to lowering the development threshold, these descriptions provide other benefits, such as reusability of existing plugins, the construction of new ones and compatibility between common output formats of different interfaces. The description-based interface generation concept is not new. For example, Javamatic (Phanouriou and Abrams, 1997) and Pise (Gilbert, 2002; no longer in use) read program information from specifically constructed XML files to output a GUI in a Java applet or static HTML form, respectively. The rapid advancement of web technologies and web browsers enables Pline to render modern, interactive GUIs as a viable alternative to standalone graphical applications. In addition, the web-standards based Pline GUIs can be seamlessly integrated into web pages without the need to install any other dependencies. Pline defines the description standard in JSON that uses minimal notation for structuring the data and is therefore more compact and easier to read and write than XML.

Although making new Pline plugins does not require programming, manually writing the JSON data fields assumes some technical knowledge and, for larger plugins, can be a tedious task. This can be addressed in a couple of ways:

1) Since the plugin JSON is a form of machine-readable program documentation, it could be converted to/from other similar formats. Instead of writing a Pline plugin from scratch, it could be programmatically created from e.g. a CWL description file, a CLI program help text or the Unix man page.

2) To make the creation of Pline plugins as simple as possible, the JSON could be constructed using a graphical web page (that itself could be made with the help of Pline).

The converters and a graphical builder tool for making Pline plugins are the subject of future work.

Pline consists of three parts: the plugin JSON API, the interface generator and the server script. These parts are independent and can be replaced with a compatible third-party module. The JSON plugin files, for example, could be used as information for generating human-readable documentation or converted to another format for use in workflow management systems. Provided with a suitable JSON file, the Pline interface generator can draw web interfaces for any use case where user input is needed. Pline interfaces communicate with the server module using standard HTTP requests. Therefore a web page integrating Pline interfaces can use its existing backend server in place of the Pline python script as long as the server accepts the command data received from the interface. The Pline server module, however, is a lightweight and installation-free implementation of a web server that can easily be bundled with the rest of Pline files and use the package as a standalone desktop application. With some modifications and using frameworks like Electron (https://electronjs.org), a Pline plugin can be converted to a native graphical desktop application with the accompanying user convenience and performance benefits.